\documentclass[preprint,aps,showpacs]{revtex4}

\usepackage{graphicx}
\usepackage{epsfig}
\usepackage[squaren]{SIunits}
\usepackage{textcomp}
\date{\today}
\begin{document}
\title {Imaging interferometry of excitons in two-dimensional structures: Can it detect exciton coherence}
\author{Heinrich Stolz}
\author{Maria Dietl}
\author{Rico Schwartz}
\author{Dirk Semkat}


\affiliation{Institut f\"ur Physik, Universit\"at Rostock,
D-18051, Rostock, Germany}
\date{\today}
\renewcommand{\baselinestretch}{1.0}
\newcommand{\be}{\begin{equation}}
\newcommand{\ee}{\end{equation}}
\newcommand{\bea}{\begin{eqnarray}}
\newcommand{\eea}{\end{eqnarray}}
\newcommand{\eps}{\varepsilon}
\newcommand{\ev}[1]{\langle {#1} \rangle}
\newcommand{\vc}[1]{\boldmath #1}
\newcommand{\ket[1]}{|{#1}\rangle}
%

\begin{abstract}
Using the theory of imaging with partially coherent light, we derive general expressions for different kinds of interferometric setups like double slit, shift and mirror interference.  We show that in all cases the interference patterns depend not only on the point spread function of the imaging setup but also strongly on the spatial emission pattern of the sample. Taking typical experimentally observed spatial emission patterns into account, we can reproduce at least qualitatively all the observed interference structures, which have been interpreted as signatures for spontaneous long range coherence of excitons, already for incoherent emitters \cite{butov2002, dubin2006}. This requires a critical reexamination of the previous work.     
\end{abstract}
%
\maketitle

Condensation of excitons is still a fascinating topic of solid state physics. 
As in every Bose-Einstein condensate, spontaneous coherence of matter waves should emerge in the exciton system. This spatial coherence is transferred to the decay luminescence and thus should be observable in the light emission from the exciton cloud. 

The standard way to measure coherence of light is by interferometry. Indeed, in the past different setups for spatial interferometry have been used to measure the spatial coherence of the light emitted by two-dimensional semiconductor structures, most often in the form of two types of interferometers: 
\begin{itemize}	
\item{Shift interferometry.} Here two images of the same object, shifted by a small vector  $\vec{\delta}$, are superimposed and by varying the phase delay between the two light paths, interference fringes are generated (see e.g. Fig.\ 5 b of Ref. \cite{kasprzak2006} and Fig.\ 2 of Ref.\ \cite{butov2012}).
To determine the interference contrast, the simple formula $C = (I_{12}-I_1-I_2)/2\sqrt{I_1 I_2}$ is used. 
\item{Mirror interferometry.} Here the second image is either the mirror or the inverted image of the object. This is obtained in a Michelson type interferometer, where the second mirror is replaced by e.g. a retro-reflector (see e.g. Fig.\ 5 c-e of Ref. \cite{kasprzak2006}).
\end{itemize}
In all the systems investigated up to now, namely exciton-polaritons in a quantum well/microcavities or excitons in double quantum well (DQW) structures, these types of interferometry are considered as the
``Standard method for detecting long range order in an exciton system'' \cite{butov2012}.
Quite remarkably, the results of all the experiments reported show a rather uniform behaviour, a pronounced increase of the coherence of the light emitted from the excitons either by lowering the temperature or by increasing the power of the laser exciting the excitons (compare Figs.\ 3 d,e, 4, and 5 of Ref.\ \cite{butov2012} or Fig.\ 5 a of Ref. \cite{kasprzak2006}).   

However, in most of the studies up to now, it has not been taken into consideration that all interference setups are also imaging systems and that one should apply the well-known theory of imaging to describe the interference phenomena properly.  Most important, in the papers one fact has been overlooked, that the coherence property of light is changed by imaging. This is based on the fundamental van Cittert-Zernike theorem of optics and well-known in the optics community. 
 
In this paper we develop a rigorous theory of imaging interferometry for an arbitrary patterned source. The properties of the interferometric setup can be desribed by an appropriate response function which is related to the amplitude point spread function \cite{bornwolf,ginmu}. The results show clearly that in addition to the point spread function also the spatial emission pattern of the sample has profound effects on the resulting interference pattern. 

The paper is organized as follows. In section 2 we derive a general theory of imaging interferometry, which we then specialise to the various setups. Explicit results are given for the case of completely incoherent emitters. In section 3 we show for some different spatial emission patterns, which are typical for the various experiments reported in the literatur, the resulting interference pattern and compare with the experiments. The paper closes with a critical discussion.  

\section{Theory of imaging interferometry}

\subsection{Theory for 2d objects}

Since the objects in this study are planar structures with thickness well below the wavelength of light, we can describe both object and image with 2d vectors. We start with a single point emitter at position $\vec{\rho}_o$ in the object plane at $-d_1$, which is imaged by a lens of focal length $f$ in the image plane at $d_2$ with $1/f=1/d_1+1/d_2$ and magnification $M=d_1/d_2$. The amplitude of the light field at a point $\vec{\rho}_i$ in the image plain at $d_2$ is then given by \cite{ginmu}
\bea
E_I(\vec{\rho}_i) & = & \frac{M}{d_1 \lambda^2}\exp\left[-i k d_1(1+1/M) \right] \cdot \exp\left[ -\frac{i k M}{2 d_1} {\vec{\rho}_i\,}^2\right]\\
 & & \times E_O(\vec{\rho}_o)\exp\left[- \frac{i k}{2 d_1} {\vec{\rho}_o\,}^2\right] \cdot P_{2d}(\vec{\rho}_o+M\vec{\rho}_i) \, ,\nonumber
\eea
with $E_O(\vec{\rho})$ denoting the field amplitude of the emitter and $P_{2d}(\vec{\rho})$ the 2d amplitude point spread function $PSF$ of the lens. 

In imaging interferometry we superimpose on this image that of an identical object but on which we impose a symmetry operation $\cal R$. This can be either a shift by a small vector $\vec{\delta}$ or a reflection, e.g. at the $yz$ plane. In addition we impose an additional phase $\Phi$ to make the interference fringes visible.  The image field of this object is given by
\bea
E'_I(\vec{\rho}_i,{\cal R})& = & \frac{M}{d_1 \lambda^2}\exp\left[-i k d_1(1+1/M)+i\Phi \right] \cdot \exp\left[ -\frac{i k M}{2 d_1} {\vec{\rho}_i\,}^2\right]\\
 & & \times E_O(\vec{\rho}_o)\exp\left[- \frac{i k}{2 d_1} ({\cal R}(\vec{\rho}_o))^2\right] \cdot P_{2d}({\cal R}(\vec{\rho}_o)+M\vec{\rho}_i)\, . \nonumber
\eea

The intensity distribution, which gives the interference pattern of a single point emitter is given by
\be \label{interference_simple}
I_{12}(\vec{\rho}_i,{\cal R}) = |E_I(\vec{\rho}_i)+E'_I(\vec{\rho}_i,{\cal R})|^2 =I_1+I_2+I_{inter} \, .
\ee
While $I_1,I_2$ are the two images of the object, the interference term is given by
\bea \label{eq:interference_point}
I_{inter} & = & 2 {\rm{Re}} \left[E_I(\vec{\rho}_i)E'^*_I(\vec{\rho}_i,\vec{\delta})\right]\\
& \propto & 2  |E_O(\vec{\rho}_o)|^2 {\rm{Re}} \Bigl\{ \exp\left[ -\frac{i k}{d_1} [ (\vec{\rho}_o)^2-({\cal R}(\vec{\rho}_o))^2]+i\Phi \right]  \\ \nonumber 
& & \times P_{2d}(\vec{\rho}_o+M\vec{\rho}_i) P^*_{2d}({\cal R}(\vec{\rho}_o)+M\vec{\rho}_i)\Bigr\} \, .\nonumber
\eea
 
To obtain the interference pattern in the most general case of many partial coherent emitters, we have to image instead of the fields the first order field correlation function $G_O(\vec{\rho},\vec{\rho}')= \langle E_O(\vec{\rho}) E^*_O(\vec{\rho}')\rangle$, which is identical to the mutual coherence function of the emitter \cite{bornwolf} by applying the van Cittert-Zernike theorem \cite{bornwolf}. This gives the following expression for the interference pattern
\bea \label{eq:interference_pattern_general}
I_{inter}(\vec{\rho}_i) &\propto & 2 \int\int  {\rm{Re}} \Bigl\{ G_O(\vec{\rho}_o,\vec{\rho}\,'_o)\exp\left[ -\frac{i k}{d_1} [ (\vec{\rho}_o)^2-({\cal R}(\vec{\rho}\,'_o))^2]+i\Phi \right]  \\ \nonumber 
& & \times P_{2d}(\vec{\rho}_o+M\vec{\rho}_i) P^*_{2d}({\cal R}(\vec{\rho}\,'_o)+M\vec{\rho}_i)\Bigr\} \nonumber
 d\vec{\rho}_o d\vec{\rho}\,'_o\,.
\eea 
Using the property 
\be
G_O(\vec{\rho},\vec{\rho}')=I_O(\vec{\rho})\delta(\vec{\rho}-\vec{\rho}') 
\ee
of an incoherent source, we see that Eq. (\ref{eq:interference_pattern_general}) goes over into 

\bea \label{eq:interference_pattern}
I_{inter}(\vec{\rho}_i) &\propto & 2 \int  I_O(\vec{\rho}_o) {\rm{Re}} \Bigl\{ \exp\left[ -\frac{i k}{d_1} [ (\vec{\rho}_o)^2-({\cal R}(\vec{\rho}_o))^2]+i\Phi \right]  \\ \nonumber 
& & \times P_{2d}(\vec{\rho}_o+M\vec{\rho}_i) P^*_{2d}({\cal R}(\vec{\rho}_o)+M\vec{\rho}_i)\Bigr\} \nonumber
 d\vec{\rho}_o\,.
\eea 
It should be noted that one obtains the same expression for the interference pattern of a totally incoherent source by integrating Eq.\ (\ref{eq:interference_point}) directly over the whole emitter, because there is no correlation between the emitting excitons.

Equations (\ref{eq:interference_pattern_general}) and (\ref{eq:interference_pattern}) are the central relations for shift interferometry of any sources. They show that the interference pattern depends not only on the PSF but also on the intensity distribution of the emitting source in a way which is not straightforward but rather complicated, a fact which has been overlooked up to now.
\section{Interferometric setups}
\subsection{Shift interferometer}
Inserting in Eq. \ref{eq:interference_pattern} the shift operation as
\be
{\cal R}(\vec{\rho})=\vec{\rho}-\vec{\delta}
\ee
and introducing the shift interferometer response function
\be\label{eq:psi_response}
P_{\rm SI}(\vec{\rho},\vec{\delta})= P_{2d}(\vec{\rho}) P^*_{2d}(\vec{\rho}-\vec{\delta}) \, ,
\ee
the different terms in Eq. (\ref{interference_simple}) can be written as a two-dimensional convolution integral between a phase shifted image and the interferometer response function
\bea \label{eq:images}
I_1(M\vec{\rho}_i) & = & I_O(\vec{\rho}) \otimes P_{\rm SI}(\vec{\rho},0)\\
I_2(M\vec{\rho}_i,M\vec{\delta}) & = & I_O(\vec{\rho}-\vec{\delta}) \otimes P_{\rm SI}(\vec{\rho},0)\\
I_{\rm inter}(M\vec{\rho},M\vec{\delta}) & = & 2 {\rm{Re}}\left\{ I_O(\vec{\rho})\exp\left(-\frac{i k}{d_1}\vec{\rho} \cdot \vec{\delta}+ i \Phi\right) \right\} \otimes P_{\rm SI}(\vec{\rho}_i,\vec{\delta}) \,.
\eea 
which allows a very efficient numerical calculation via fast Fourier transform.
\subsection{Mirror interferometry}
In case of inversion interferometry, the symmetry operation is given as
\be
{\cal R}(\vec{\rho})=-\vec{\rho}
\ee
and Eq. (\ref{eq:interference_pattern}) goes over to
\bea \label{eq:interference_pattern_mirror}
I_{inter}(\vec{\rho}_i) &\propto & 2 \int  I_O(\vec{\rho}_o) {\rm{Re}}\Bigl\{ \exp\left[i\Phi \right]  \\ \nonumber 
& &  P_{2d}(\vec{\rho}_o+M\vec{\rho}_i) P^*_{2d}((-\vec{\rho}_o)+M\vec{\rho}_i)\Bigr\} \nonumber
 d\vec{\rho}_o\,.
\eea 
which, however, is not a simple convolution, making the calculation a bit more tedious. 
\section{Point spread functions}
\subsection{Standard setup}
\begin{figure}[tbh]
 \begin{center}
 \begin{minipage}{8cm}
  \includegraphics[width=\textwidth]{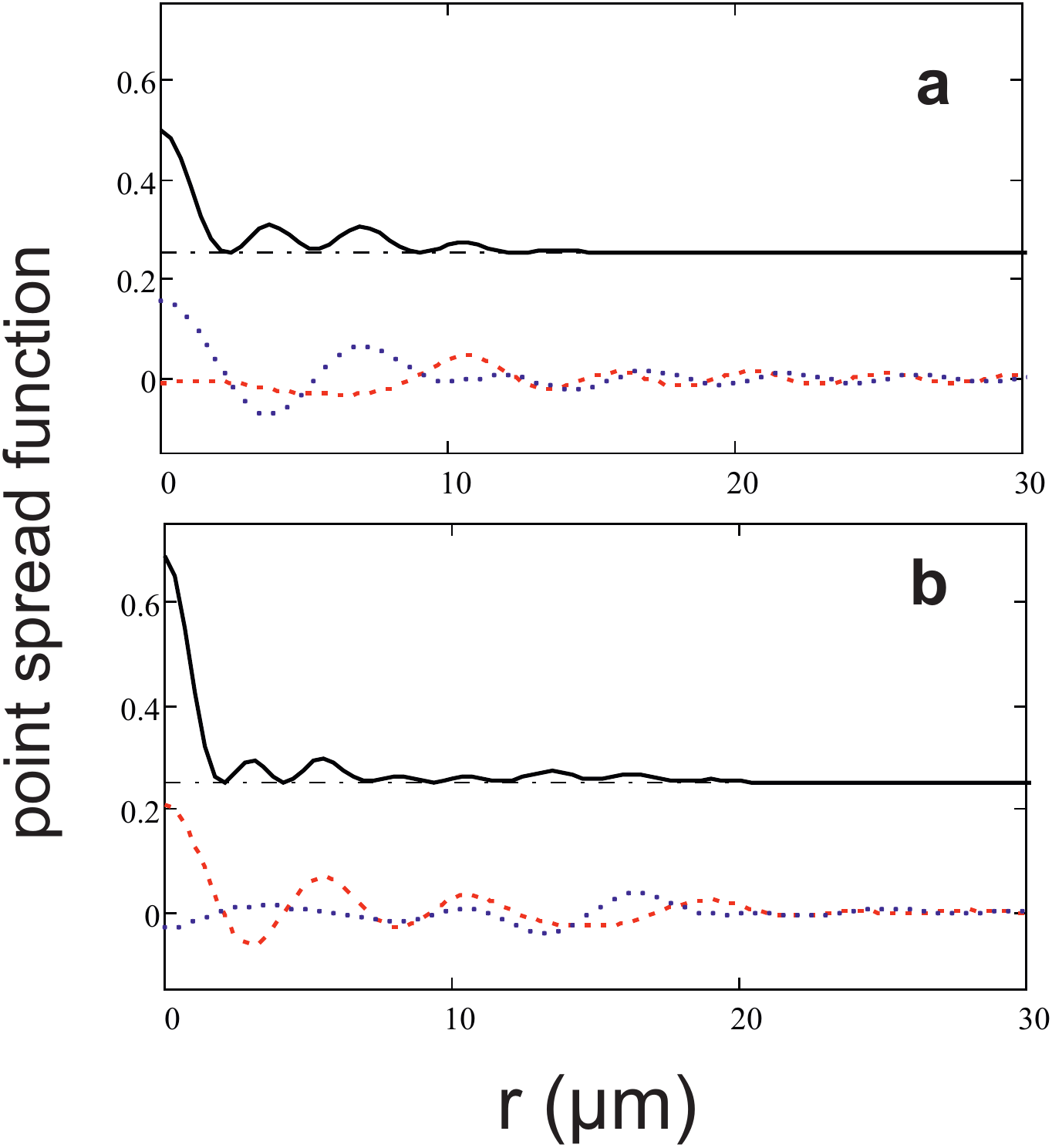}
 \end{minipage}
 \caption{Two-dimensional sections through the 3d point spread function of the standard imaging setup for different defocussing. Panel a) $z=0$ and panel b) $z=-9.2 \mu$m. The real part is given by the dashed line, the imaginary part by the dotted line and the absolute square by the full line (note shift of zero of the $y$ axis).}
 \label{fig:psf3d}
 \end{center}
\end{figure}
Since the interference pattern depends on the (amplitude) point spread function of the optical setup, we first discuss two typical cases. The first one is the standard setup used in most experiments up to now \cite{kasprzak2006, deng2007}. Here, the sample is mounted inside an optical cryostat onto a cold finger which cools the sample to liquid Helium temperatures. Optical access is through a quartz window with thickness of typical 1 to 2 mm. Imaging is performed with a high numerical aperture microscope objective with e.g. $N.A. = 0.5$ and long working distance. Assuming the microscope objective to be perfectly corrected for optical aberrations, the window still gives rise to quite severe spherical abberations. The three dimensional point spread function of the whole setup, from which we obtain the 2d part by setting $z$ in the optimal focus, can be approximated quite well by that of a thin lens with sperical aberrations in the paraxial limit \cite{bornwolf,ginmu}:  
\be\label{eq:psf1}
P_{3d}(r,z)=P_0 \int_0^1 \rho \exp \left( i \left(\frac{a}{d_1} \right)^2 k z \right) \exp \left( - i k \Phi (\rho) \right) J_0(k a/d_1 r \rho) d \rho \,.
\ee
with $\rho$ the radial distance from the optical axis in unit of the lens radius $a$ and $J_0(x)$ the Bessel function of zeroth order.
The wavefront aberration $\Phi$ is given by
\be \label{eq:wave1}
\Phi(\rho)= \frac{1}{\sqrt{2}} A_{040}R^0_4(\rho)
\ee
with $R^0_4(\rho)$ denoting Zernike's circle polynomial of order $n=4,l=0$ and $A_{040}$ the Zernike coefficient for primary spherical aberration. 

For a plane parallel plate of thickness D and index of refraction $n_P$ this is given by \cite{wynad}
\be
A_{040}= -\sqrt{2}\frac{D (n_p^2-1)}{48 n_P^3}(\arcsin(N.A.))^4 
\ee 
For typical values $D=1$~mm and $n_P=1.45$ we have $A_{040}=-0.8\, \micro\rm m$, which is of the order of the wavelength and thus not small. 
The 3d PSF for this situation obtained by numerical integration of Eq. \ref{eq:psf1} is shown in figure \ref{fig:psf3d} for different defocusing. While for the optimum defocussing conditions the apparent resolution given by the FWHM of the PSF is still quite good, strong coherent side lobes extend quite far out of the center. As schown below this results in such sever distortions of the interference pattern making a simple interpretation of experimental results almost impossible.
\subsection{Optimized setup}

\begin{figure}[hbt]
 \begin{center}
 \begin{minipage}{14cm}
  \includegraphics[width=\textwidth]{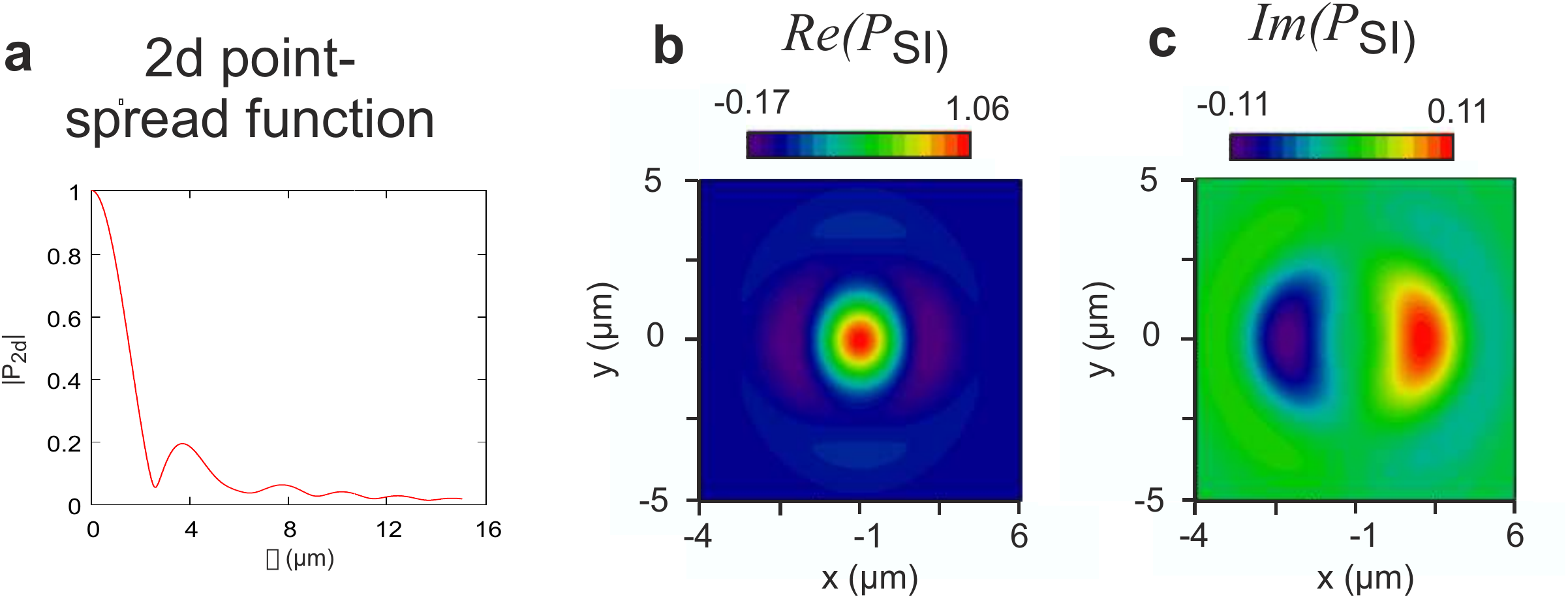}
 \end{minipage}
 \caption{Two-dimensional point spread function (Panel a), and real and imaginary parts of the corresponding shift interferometer response function (panel b and c). }
 \label{fig:psf1}
 \end{center}
\end{figure}
Much less aberrations are introduced, when the imaging lens is inside the cryostate, directly facing the sample, as was the case in  Ref.\ \cite{butov2012a}. Here even the 
PSF of the actual imaging setup can be obtained from Fig.\ 3a. 
For our purposes, we also approximate the PSF by that of a thin lens with sperical aberrations. Chosing $a/d_1=0.175$ and $A_{040}=0.191 \lambda$ gives a reasonable fit (compare Fig. \ref{fig:psf1}a with Figure 3a of Ref. \cite{butov2012a}).
In panel b and c of Fig. \ref{fig:psf1}, we show an example for the shift interferometer response $P_{\rm SI}$ for $\delta = 2\,\micro\rm m$. It shows that $P_{\rm SI}$ possesses a substantial imaginary part, which will give rise to highly structured spatial interference patterns (see Fig. 2 and 3).

\section{Results}
To demonstrate the essential points of our argumentation, we first discuss in the following two examples of intensity patterns which are typical for the observation in Ref. \cite{butov2012a}. The first one is a ring-type structure which is similar to an LBS ring (compare Fig. 1a and Fig. 1S of the supplementary information of Ref. \cite{butov2012a}) and may be given as $I_O(\rho)\propto \exp[-(\rho-\rho_R)^2/\sigma_R^2]$. The size of the ring was chosen to be $2 \rho_R=4.9\, \micro\rm m$ diameter and $\sigma_R=2 \micro\rm m$ resulting in a small dip in the middle (see Fig. 2a). For the additional phase we choose $\Phi(y)=\pi/2 y$ to reproduce the experimental interference fringes. The spatial image of the interference pattern is shown in Fig. 2b, the pattern of interference contrast is given in panel c. The two images show the same pattern as found in the experiment, low contrast in the spot center, but a high contrast which reaches almost 1 in a left and right side lobe of the spot. As can be seen in panel b, we even find fork-like interference patterns at the boundary of the luminescing spot. It has to be stressed, that all these signatures arise already for a {\it completely incoherent} emitting source!

\begin{figure}[h]
 \begin{center}
 \begin{minipage}{12cm}
  \includegraphics[width=\textwidth]{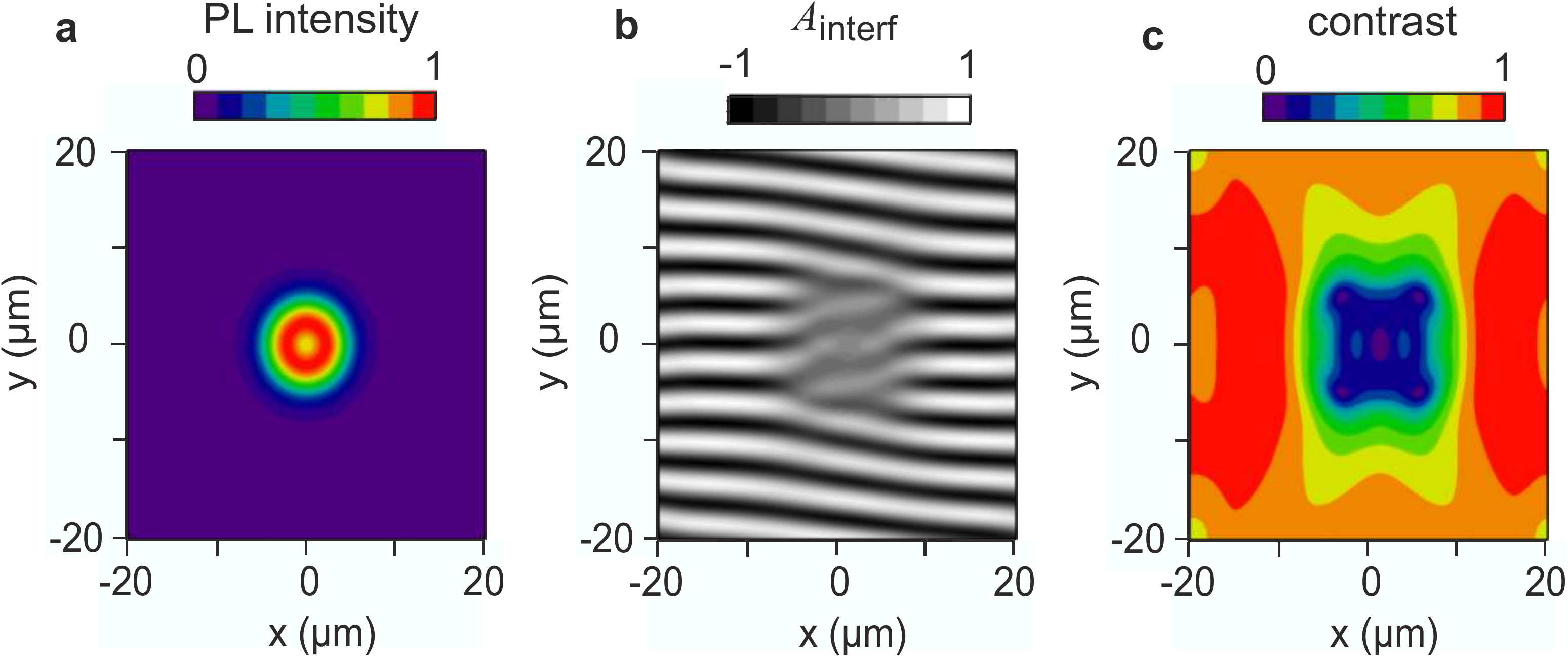}
 \end{minipage}
 \caption{Luminescence pattern of an incoherently emitting ring (panel a), the interference contrast calculated by Eq. \ref{eq:interference_pattern} (panel b) and the interference contrast (panel c).}
 \label{fig:contrast_lbs}
 \end{center}
\end{figure}

\begin{figure}[h]
 \begin{center}
 \begin{minipage}{12cm}
  \includegraphics[width=\textwidth]{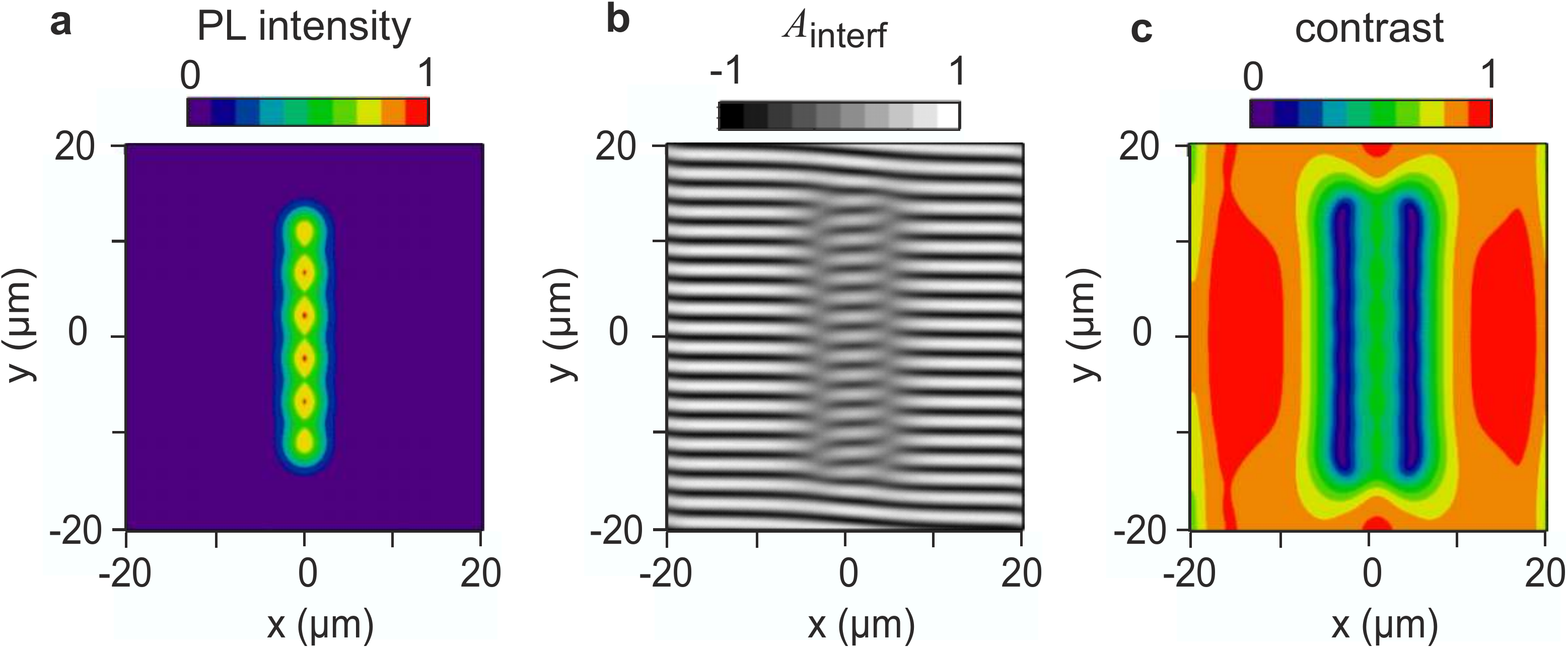}
 \end{minipage}
 \caption{Luminescence pattern of a chain of incoherent emitters (panel a), the interference contrast calculated by Eq. \ref{eq:interference_pattern} (panel b) and the interference contrast (panel c).}
 \label{fig:contrast_outr}
 \end{center}
\end{figure}

As second example we consider the interference pattern of a line array of spots, similar to those in the ``macroscopically ordered exciton state'' of an outer ring (see Fig.2a).    
Each spot is represented by a Gaussian as $I_O(\rho)\propto \exp[-(\rho/\sigma_O)^2 ] $
with $\sigma_O=2\, \micro\rm m$, the distance of the spots being $4.5\, \micro\rm m$. The interference pattern and contrast for such a source is shown in panels b and c. Again, we reproduce almost quantitavely the essential observations of Ref. \cite{butov2012a}, even the two parallel lines of minimum contrast can be identified.

\begin{figure}[h]
 \begin{center}
 \begin{minipage}{12cm}
  \includegraphics[width=\textwidth]{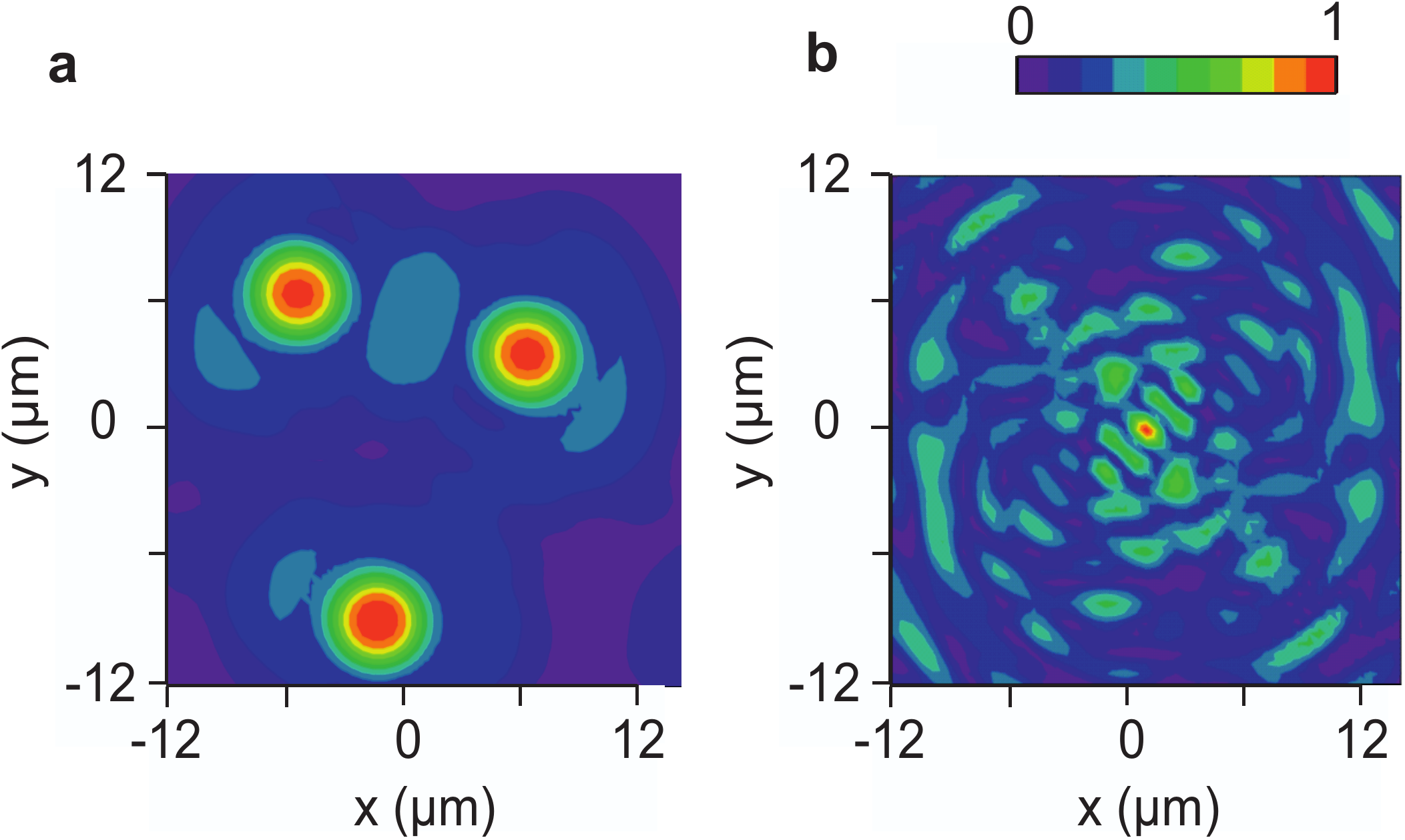}
 \end{minipage}
 \caption{Luminescence pattern of an array of three random incoherent emitters (panel a) and the interference contrast of a mirror type interferometer setup calculated with the point spread function of Fig. 1 for a standard optical setup (panel c).}
 \label{fig:mirror_interference}
 \end{center}
\end{figure}

Finally, we show results for the case of mirror interferometry, which e.g. has been used in Ref. \cite{kasprzak2006} to claim spontaneous coherence in a dense polariton system. 
Here the authors report that the large emitting spot breaks down to an array of small spots with sizes in the range of $2 - 3 \mu$m (see Fig. 4 panel f and h of Ref. \cite{kasprzak2006}). We therefore simulated the mirror interference pattern by positioning small incoherent emitters of such sizes in the object plane (see Fig. \ref{fig:mirror_interference} panel a). The resulting contrast image is shown in Fig. \ref{fig:mirror_interference} panel b. The pattern is very similar to that observed experimentally, even the maximum amount of contrast (30\%) is identical to that found in the experiments.

\section{Conclusions}
In conclusion, it is obvious from our model calculations that the experimental findings upon which the claims of detecting spontaneous coherence of excitons and exciton polaritons in Ref. \cite{butov2012a,butov2012} and \cite{kasprzak2006} are based, can be explained straightforwardly by the properties of partial coherent light without the assumption of exciton coherence, by taking only into account the spatial emission patterns characteristic of the samples. Indeed, all these patterns have in common that one observes a change in the pattern from a spatially rather large and homegeneous distribution to one with an array of small spots with sizes approaching the limit of optical resolution of the imaging setup. The observation of optical coherence in this case is not surprising, since in the limit of a point source, the emitted light is by definition completely coherent. Therefore, the results of the above mentioned papers have to be reconsidered by taking the effects of imaging properly into account before any claims to have observed spontaneous coherence in an exciton system can be justified,. 

Generally, we want to state that using interferometric methods for determining coherence of exciton systems is highly questionable. Especially, arguments that are not based on a rigorous theoretical analysis may turn out to be completely misleading.



\end{document}